\documentclass[prb,nofootinbib,twocolumn,superscriptaddress]{revtex4} 

\usepackage{graphicx}
\usepackage{dcolumn}
\usepackage{bm}
\usepackage{threeparttable}
\usepackage{times}
\usepackage{mathptmx}
\usepackage{lscape}
\usepackage{natbib}
\usepackage{amsmath}
\usepackage{amssymb}
\usepackage{braket}
\usepackage{comment}
\usepackage{color}
\usepackage{float}

\def\degree{\kern-.2em\r{}\kern-.3em}

\begin{document}


\title{  Exact Determination of Moments for Density of States in Multidimensional Configuration Space  }

\author{Shouno Ohta}
\affiliation{
Department of Materials Science and Engineering,  Kyoto University, Sakyo, Kyoto 606-8501, Japan\\
}%

\author{Koretaka Yuge}
\affiliation{
Department of Materials Science and Engineering,  Kyoto University, Sakyo, Kyoto 606-8501, Japan\\
}%

\begin{abstract}
{
For classical discrete systems on periodic lattice under constant composition $x$, we derive explicit expression of any-order moments for configurational density of states (CDOS). 
The derived expression clarifies that any-order moments can always be given by linear combination of the first-order moments, whose coefficient 
depends on geometric information of lattice. The expression enables us to \textit{exactly} determine system-size ($N$) dependence of moments, 
where analytic representation in terms of $N$ and $x$ is explicitly given up to lower-order generalized moment. Validity of the derived expression is confirmed by exact estimation of moments for binary system bcc with finite system size, considering all possible atomic configuration. 
  }
\end{abstract}


\maketitle

\section{Introduction}
For classical discrete system under constant composition, expectation value of physical quantity $A$ can be typically obtained through the so-called canonical average, 
\begin{eqnarray}
\Braket{A} = Z^{-1}\sum_i A_i \exp{\left(-\beta E_i\right)},
\end{eqnarray}
where summation is taken over possible microscopic states on configuration space, 
$Z$ denotes partition function 
\begin{eqnarray}
Z = \sum_i \exp{\left(-\beta E_i\right)},
\end{eqnarray}
which plays central role in statistical mechanics, since it directly determine Helmholtz free energy through $F=-kT\ln Z$.
When we introduce basis functions $\left\{q_1,\cdots, q_f\right\}$ for $f$-dimensional configuration space, $Z$ can be rewritten as multiple integral:
\begin{eqnarray}
Z =\mathop{\int\cdots\int} g\left(q_1, \cdots, q_f\right) \exp{\left(-\beta \sum_\alpha \Braket{q_\alpha |E}q_\alpha\right)} dQ,
\end{eqnarray}
where $g$ denotes CDOS, and $\Braket{\quad|\quad}$ represents inner product, i.e., trace over configuration space. 
The equation indicates that when information about multibody interactions (corresponding to specify the values of inner products) are once given, $Z$ can be exactly determined \textit{if} one can fully know
the landscape of CDOS, $g\left(q_1, \cdots, q_f\right)$.
However, so far, exact landscape for even one-dimensional CDOS is not clarified for three dimensional lattice, or even moments of one-dimensional CDOS and their system-size dependence is not quantitatively elucidated. 
Therefore, alternative approaches has been amply developed to effectively sample microscopic states dominantly contributing to macroscopic properties, including Metropolis algorithm, entropic sampling and Wang-Landau sampling.\cite{mc1,mc2,mc3,wl} 
Here we derive explicit expression of any-order moments (and their combinations) of cluster correlation function $\left\{\xi_1,\cdots, \xi_f\right\}$ in cluster expansion, and clarifies that any moments (and their combinations) can be given by linear combination of the first-order moments, whose coefficient depends on the number of specific figures included in given lattice. 
The present approach is the partial extention of our previous approach, where CDOS is found to asymptotically get close to multidimensional gaussian in terms of macroscopic as well as microscopic viewpoints.\cite{em1,em3,cm,em2}
The expression enables us to exact determination of moments, 
where analytic representation in terms of $N$ and $x$ is given up to lower-order generalizedmoment. Validity of the derived expression is confirmed by exact estimation of moments for finite-size system on fcc and bcc lattice. The details are shown below.

\section{Derivation and Applications}
In order to see the basic concept for the present derivation of any-order moments, let us first derive exact expression of the first-order moment of CDOS for any given cluster.
Consider A$_{1-x}$B$x$ binary system with $N$ lattice points, focusing on $m$-body figure $c$. Then corresponding first-order moment of the correlation function $\Braket{\xi_c}$ is first given by
\begin{equation}
	\langle\xi_c\rangle = \dfrac{1}{{}_{N}{\rm C}_{xN}}\displaystyle\sum_{\sigma}\xi_r(\sigma) = \dfrac{1}{{}_{N}{\rm C}_{xN}}\displaystyle\sum_{\sigma}\xi_{r(1)}(\sigma)	,
\end{equation}
where summation is taken over all possible microscopic states (i.e., atomic configurations) $\sigma$, $ \xi_{c(1)} $ denotes  the correlation function of one of the symmetry-equivalent figure to $c$, $c\left(1\right)$. 
When we rewrite the correlation function in terms of the number of configuraions where B atoms occupy the figure $c\left(1\right)$, we get
\begin{eqnarray}
\label{eq:1dm}
\langle\xi_c\rangle = \dfrac{1}{{}_{N}{\rm C}_{xN}}\displaystyle\sum_{k=0}^{m}(-1)^k\cdot{}_{m}{\rm C}_{k}\cdot{}_{N-m}{\rm C}_{xN-k},
\end{eqnarray}
where $k$ denotes the number of B atoms in figure $c\left(1\right)$, ${}_{m}{\rm C}_{k}$ denotes the number of cases where $m$ lattice points of $c\left(1\right)$ are occupied by $m-k$ A and $k$ B atoms, and the rest term ${}_{N-m}{\rm C}_{xN-k}$ of cases where rest lattice points by rest atoms. 
\begin{figure}[H]
	\begin{center}
		\includegraphics[width=0.94\linewidth]
		{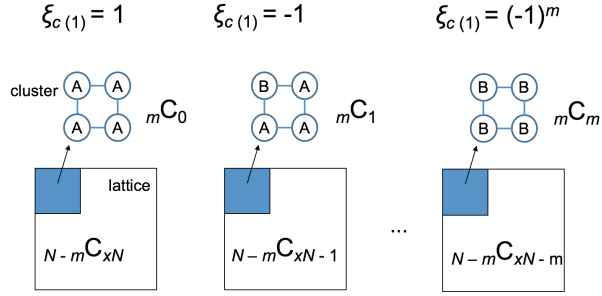}
		\caption{Schematic illustration of the concept for derivation of first-order moment of CDOS, in terms of the number of cases for the considered figure.}
		\label{fig:1}
	\end{center}
\end{figure}
\noindent Equation~(\ref{eq:1dm}) can be immediately rewritten by the following equation using simple deformation 
\begin{eqnarray}
\label{eq:1dm-2}
\langle\xi_c\rangle = \dfrac{1}{{}_{N}{\rm C}_{m}}\displaystyle\sum_{k=0}^{m}(-1)^k\cdot{}_{(1-x)N}{\rm C}_{m-k}\cdot{}_{xN}{\rm C}_{k},
\end{eqnarray}
which corresponds to explicitly considering the number of cases where individual atoms occupies each lattice point in figure $c\left(1\right)$. 
Figure~\ref{fig:1} shows schematic illustration of the concept of above equations, in terms of the number of cases for the considered figure.

When we focus on equiatomic composition, Eq.~(\ref{eq:1dm-2}) can be simplified to
\begin{widetext}
\begin{eqnarray}
\label{eq:1dm-0.5}
\langle\xi_c\rangle = \dfrac{1}{{}_{N}{\rm C}_{m}}\displaystyle\sum_{k=0}^{m}(-1)^k\cdot{}_{\frac{N}{2}}{\rm C}_{m-k}\cdot{}_{\frac{N}{2}}{\rm C}_{k}=
\begin{cases}
\dfrac{(-1)^{\frac{m}{2}}\cdot{}_{\frac{N}{2}}{\rm C}_{\frac{m}{2}}}{{}_{N}{\rm C}_{m}} & (m~:~{\rm even})\\
0 & (m ~:~{\rm odd}).
\end{cases}
\end{eqnarray}
\end{widetext}
To derive the above equation, we consider coefficient comparison of the following identity about $y$ based on Vandermonde convolution:
\begin{eqnarray}
(1-y^2)^{\frac{N}{2}} = (1-y)^{\frac{N}{2}}\cdot(1+y)^{\frac{N}{2}}.
\end{eqnarray}
Then we finally obtain exact expression for the first-order moment of $\xi_c$ at equiatomic composition, namely,
\begin{eqnarray}
\langle\xi_c\rangle = 
\begin{cases}
\dfrac{(-1)^m\cdot(m-1)!!}{(N-1)(N-3)\cdots(N-(m-1))} & (m~:~{\rm even})\\
0 & (m ~:~{\rm odd}).
\end{cases}\label{eq:1dm-0.5-2}
\end{eqnarray}
In the followings, explicit expression for higher-order moments are derived based on the above concepts for first-order moment.

\begin{figure*}[ht]
	\centering
	\includegraphics[width=0.8\linewidth]{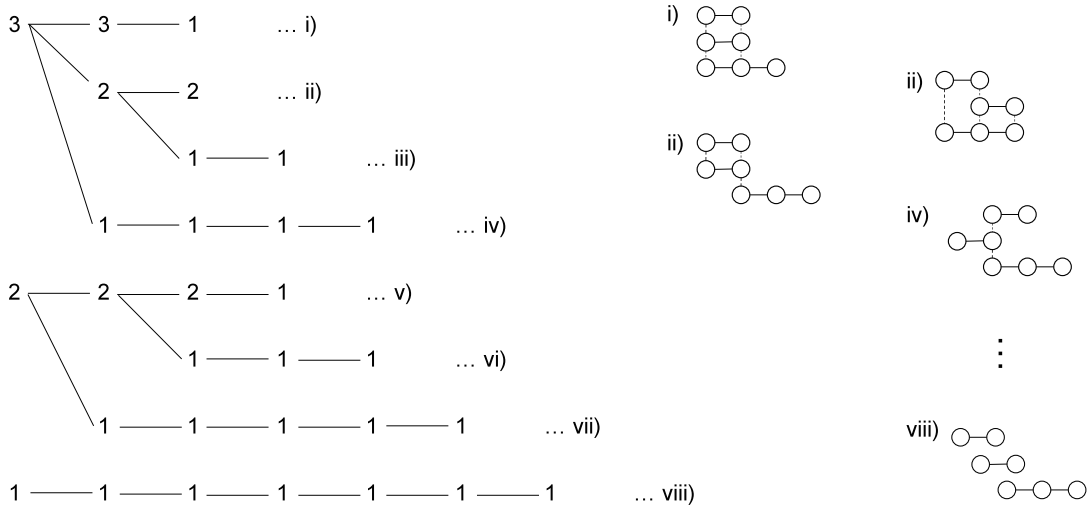}
	\caption{
	\label{fig:cluster}The corresponding between dividing natural number, 7 and example of how two pair figures and one triplet figure occupy the lattice points. The vertical dotted lines express that two or more figures occupy one lattice point.}
\end{figure*}

\begin{figure}[H]
	\centering
	\includegraphics[width=0.93\linewidth]{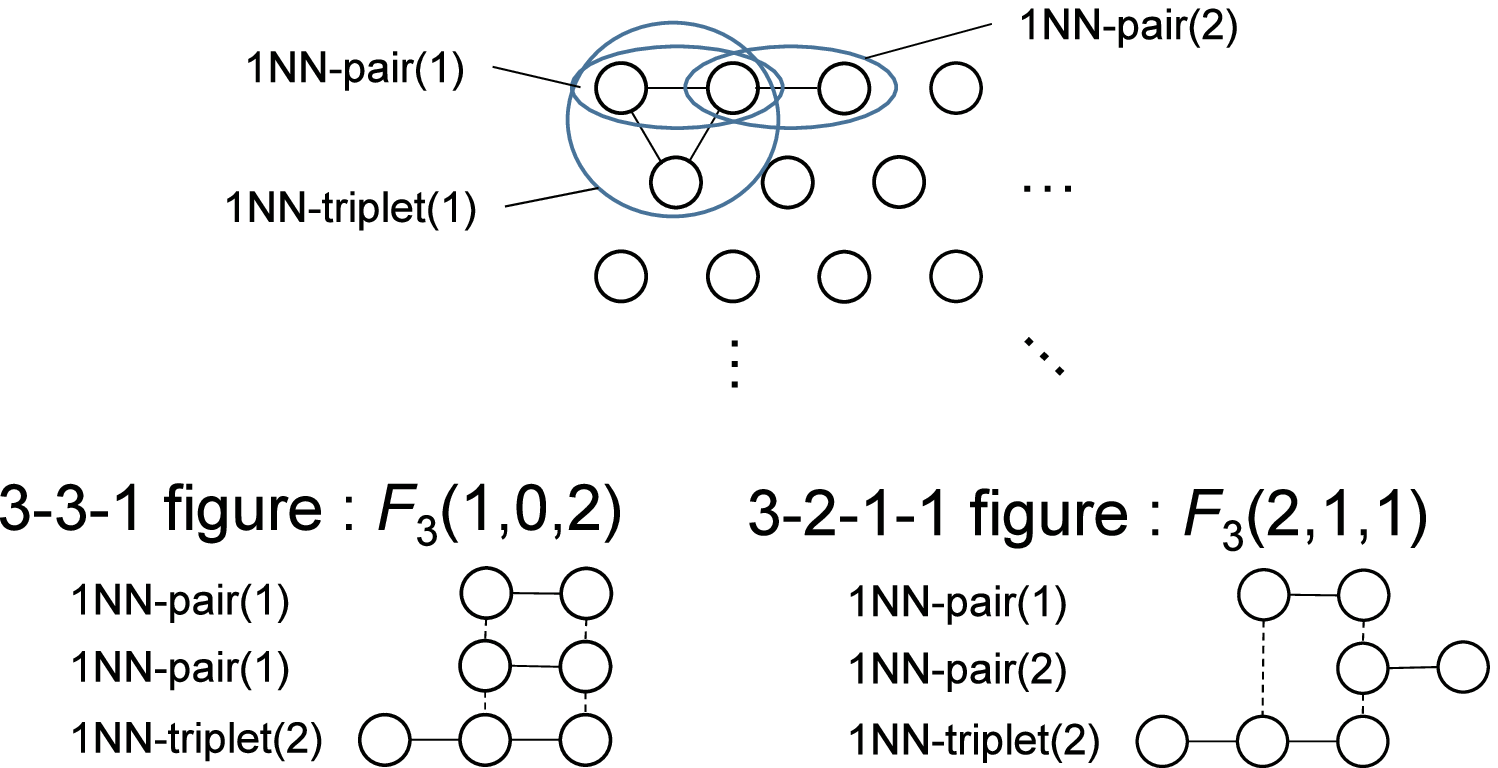}
	\caption{
	\label{fig:cluster2}Schematic illustration of corresponding between the way of occupying lattice points and the products of correlation function.}
\end{figure}

We first consider generalized moment $ \langle\xi_{c_1}^{r_1}\xi_{c_2}^{r_2}\cdots\xi_{c_n}^{r_n}\rangle $ for correlation functions $ \{\xi_{c_1},~\xi_{c_2},\cdots,~\xi_{c_n}\} $, i.e.,  combination of $ \{m_1,~m_2,\cdots,~m_n\} $-body figure $ \{c_1,~c_2,\cdots,~c_n\} $ by definition,
\begin{widetext}
\begin{eqnarray}
\label{eq:mdm}
	\Braket{\xi_{c_1}^{r_1}\xi_{c_2}^{r_2}\cdots\xi_{c_n}^{r_n}} &=& \dfrac{1}{{}_{N}{\rm C}_{xN}}\displaystyle\sum_{\sigma}\xi_{c_1}^{r_1}\xi_{c_2}^{r_2}\cdots\xi_{c_n}^{r_n} \nonumber \\
	& =& \dfrac{1}{{}_{N}{\rm C}_{xN}}\displaystyle\sum_{\sigma}(\dfrac{\xi_{c_1(1)} + \cdots +\xi_{c_1(N_{c_1})}}{N_{c_1}})^{r_1}
	\cdots(\dfrac{\xi_{c_2(1)} + \cdots +\xi_{c_n(N_{c_n})}}{N_{c_n}})^{r_n}\nonumber \\
	& =& \dfrac{1}{{}_{N}{\rm C}_{xN}\cdot N_{c_1}^{r_1}\cdots N_{c_n}^{r_n}}\displaystyle\sum_{\sigma}({\xi_{c_1(1)} + \cdots +\xi_{c_1(N_{c_1})}})^{r_1}
	\cdots({\xi_{c_2(1)} + \cdots +\xi_{c_n(N_{c_n})}})^{r_n},
\end{eqnarray}
\end{widetext}
where $N_{c_k}$ is the number of figure $ c_k $ in the lattice considered. The terms in the summation of Eq.~(\ref{eq:mdm}) is classified in terms of how the figures occupy the lattice points; more concretely, the number of lattice points occupied by individual figures and the number that each of the lattice points occupied by the figures. Mathematically, this corresponds to dividing natural number  $ m_1r_1 + \cdots m_{n}r_n $ into $ r_1 + \cdots r_n $ or less natural numbers. For example, for moment of the product of a squared pair correlation function and a triplet correlation function , $\Braket{\xi_{\rm pair}^{2} \xi_{\rm tri}}$, Figure \ref{fig:cluster} shows the correspondence between dividing natural number $2\times 2+3=7$ and how two pair figures and one triplret figure occupy the lattice points.
Let us define that $ F_{r_1 + \cdots +r_n}(p_1,~p_2,\cdots,p_{r_1 + \cdots + r_n}) $ is the $\sum_i r_i$-variable function independent of atomic configuration, which provides the number of possible \textit{diagrams} in the system characterized by its arguments: Here, $\alpha$-th argument of $F$ denotes the number of lattice points $\alpha$-times shared in the considered single diagram, including $r_1$ $c_1$-figure, $r_2$ $c_2$-figure, $\cdots$, $r_n$ $c_n$ figures. 
$f(p_1,~p_2,\cdots,p_{r_1 + \cdots + r_n})$ as the summation of spin product for a (representative, any) \textit{single} diagram $(p_1,~p_2,\cdots,p_{r_1 + \cdots + r_n})$ over all possible configurations. 
Then Eq.~(\ref{eq:dmd}) can be exactly rewritten as 
\begin{widetext}
\begin{eqnarray}
\Braket{\xi_{c_1}^{r_1}\xi_{c_2}^{r_2}\cdots\xi_{c_n}^{r_n}}\hspace{135mm}\nonumber\\ 
= 
	\dfrac{1}{{}_{N}{\rm C}_{xN}\cdot N_{c_1}^{r_1}\cdots N_{c_m}^{r_m}}\{ 
	F_{r_1 + \cdots +r_n}(m_1r_1 + \cdots m_{n}r_n,~0,\cdots,~0)\cdot f_{r_1 + \cdots +r_n}(m_1r_1 + \cdots m_{n}r_n,~0,\cdots,~0)\hspace{15mm}\nonumber\\
	 + F_{r_1 + \cdots +r_n}(m_1r_1 + \cdots m_{n}r_n - 2,~1,~0,\cdots,~0)\cdot f_{r_1 + \cdots +r_n}(m_1r_1 + \cdots m_{n}r_n - 2,~1,~0,\cdots,~0)+\cdots\}.\hspace{0mm}\label{eq:dmd}
\end{eqnarray}
\end{widetext}
When we apply Eq.~(\ref{eq:pair2tri3}) to the moment $\Braket{\xi_{\rm pair}^{2} \xi_{\rm tri}}$, we get
\begin{widetext}
\begin{eqnarray}
\Braket{\xi_{\rm pair}^{2}\xi_{\rm tri}}& =& \dfrac{1}{{}_{N}{\rm C}_{xN}\cdot N_{\rm pair}^{2}\cdot N_{\rm tri}}\displaystyle\sum_{\sigma}({\xi_{\rm pair(1)} + \xi_{\rm pair(2)}}+\cdots)^{2}(\xi_{\rm tri(1)} +\xi_{\rm tri(2)}+\cdots)\nonumber\\
&=&\dfrac{1}{{}_{N}{\rm C}_{xN}\cdot N_{\rm pair}^{2}\cdot N_{\rm tri}}\displaystyle\sum_{\sigma}(\underset{{\rm counted~as~3-3-1 ~figure }}{\underline{\xi_{\rm pair(1)}\xi_{\rm pair(1)}\xi_{\rm tri(1)}}} + \underset{{\rm counted~as~ 3-2-1-1 ~figure}}{\underline{\xi_{\rm pair(1)}\xi_{\rm pair(2)}\xi_{\rm tri(1)}}} + \cdots)\nonumber\\
&=&\dfrac{1}{{}_{N}{\rm C}_{xN}\cdot N_{\rm pair}^{2}\cdot N_{\rm tri}}(F_3(7,0,0)f_3(7,0,0)+F_3(5,1,0)f_3(5,1,0)+F_3(3,2,0)f_3(3,2,0)+F_3(1,3,0)\nonumber\\&&\hspace{0mm}f_3(1,3,0)+F_3(4,0,1)f_3(4,0,1)+\underline{F_3(2,1,1)}f_3(2,1,1)+F_3(0,2,1)f_3(0,2,1)+\underline{F_3(1,0,2)}f_3(1,0,2)). \label{eq:pair2tri3}
\end{eqnarray}
\end{widetext} 
For example, intuitive interpretation of $ F_3(2,1,1) $ and $ F_3(1,0,2) $ in Eq.~(\ref{eq:pair2tri3}) are given in Figure \ref{fig:cluster2}.
Here, for spin products summed over all possible atomic arrangements, all odd-times occupation is reduced to a single-time occupation, and even-times occupation can be diminished zero-times occupation due to the definition of spin variable, $\sigma=\pm 1$. This immediately leads to our final, general expression of the generalized moments:
\begin{widetext}
	\begin{eqnarray}
	\label{eq:mdm3}
	\langle\xi_{c_1}^{r_1}\xi_{c_2}^{r_2}\cdots\xi_{c_n}^{r_n}\rangle =	\dfrac{1}{{}_{N}{\rm C}_{xN}\cdot N_{c_1}^{r_1}\cdots N_{c_m}^{r_m}}(F'f_1(0) + F''f_1(1) + \cdots + F^{( m_1r_1 + \cdots m_{n}r_n)}f_1(m_1r_1 + \cdots m_{n}r_n)),
	\end{eqnarray}
\end{widetext}
where 
\begin{eqnarray}
F^{\left(r\right)} &=& \sum_{d_1,\cdots,d_{r_1+\cdots + r_n}}F\left(d_1,\cdots,d_{r_1+\cdots +r_n}\right) \nonumber \\
r&=&\sum_{i\in \textrm{odd}} d_i
\end{eqnarray}
$ f_1(k)/{}_{N}{\rm C}_{xN} $ in Eq.~(\ref{eq:mdm3}) is equivalent to the first-order moment derived above, which directly means that any generalized moments on multidimensional CDOS can be exactly expressed by linear combination of the first-order moments. According to this logic Eq.~(\ref{eq:pair2tri3}) can be rewritten as
\begin{widetext}
\begin{eqnarray}
\Braket{\xi_{\rm pair}^{2}\xi_{\rm tri}}
&=&\dfrac{1}{{}_{N}{\rm C}_{xN}\cdot N_{\rm pair}^{2}\cdot N_{\rm tri}}(F'f_1(1)+F''f_1(3)+F'''f_1(5)+F''''f_1(7))\nonumber
\\
&=&\dfrac{1}{{N_{\rm pair}^{2}\cdot N_{\rm tri}}}(F'\Braket{\xi_1}+F''\Braket{\xi_3}+F'''\Braket{\xi_5}+F''''\Braket{\xi_7}). \label{eq:pair2tri3-2}
\end{eqnarray}
\end{widetext} 
, in which $ \{\xi_1,~\xi_3,~\xi_5,~\xi_7\}$ are first-order moment of any $ \{1,~3,~5,~7\} $-body cluster correlation function  respectively.
\begin{figure}[H]
	\centering
	\includegraphics[width=0.93\linewidth]{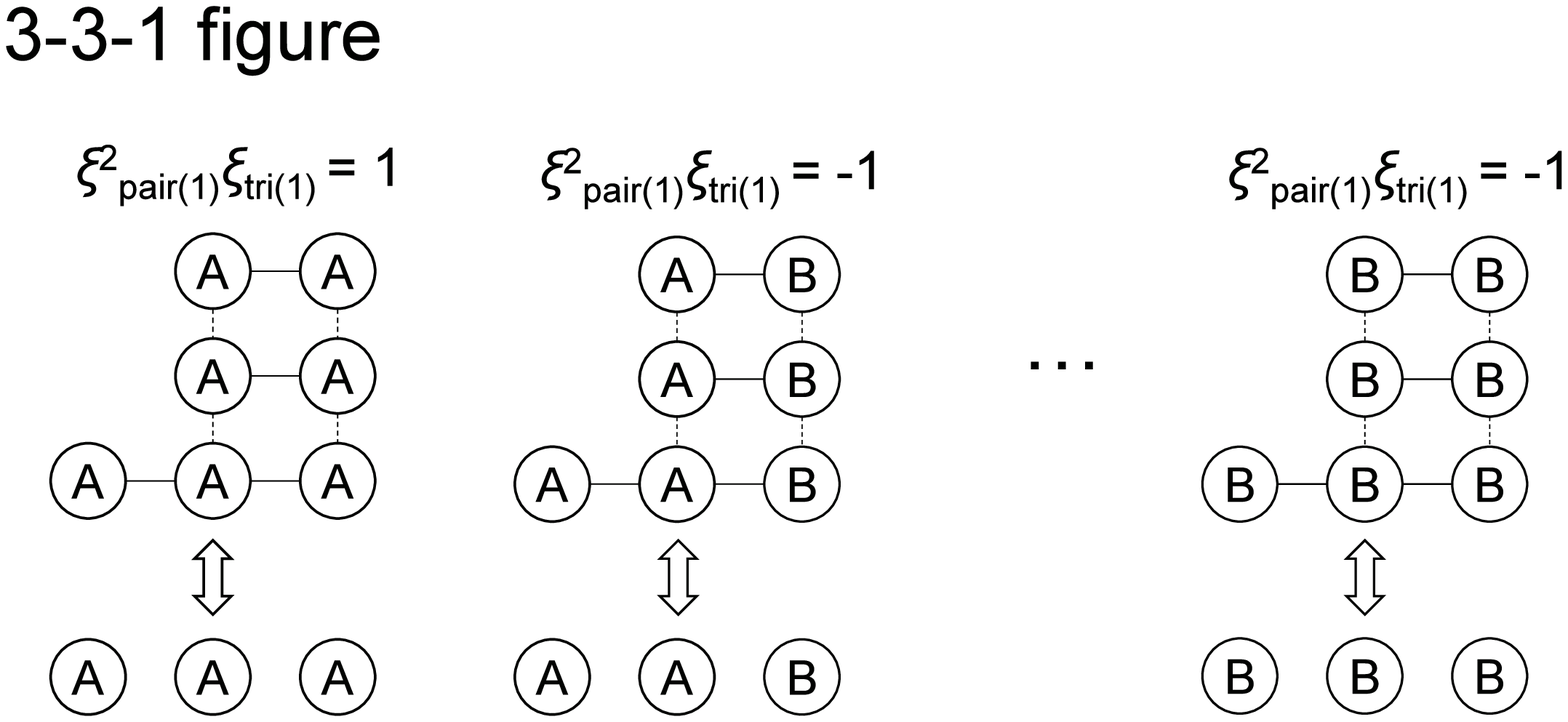}
	\caption{
		\label{fig:cluster331}Schematic illustration that shows 3-3-1 figure can be considered as 1-1-1 figure deu to the definition of spin variable.}
\end{figure}
\begin{figure}[H]
	\centering
	\includegraphics[width=0.93\linewidth]{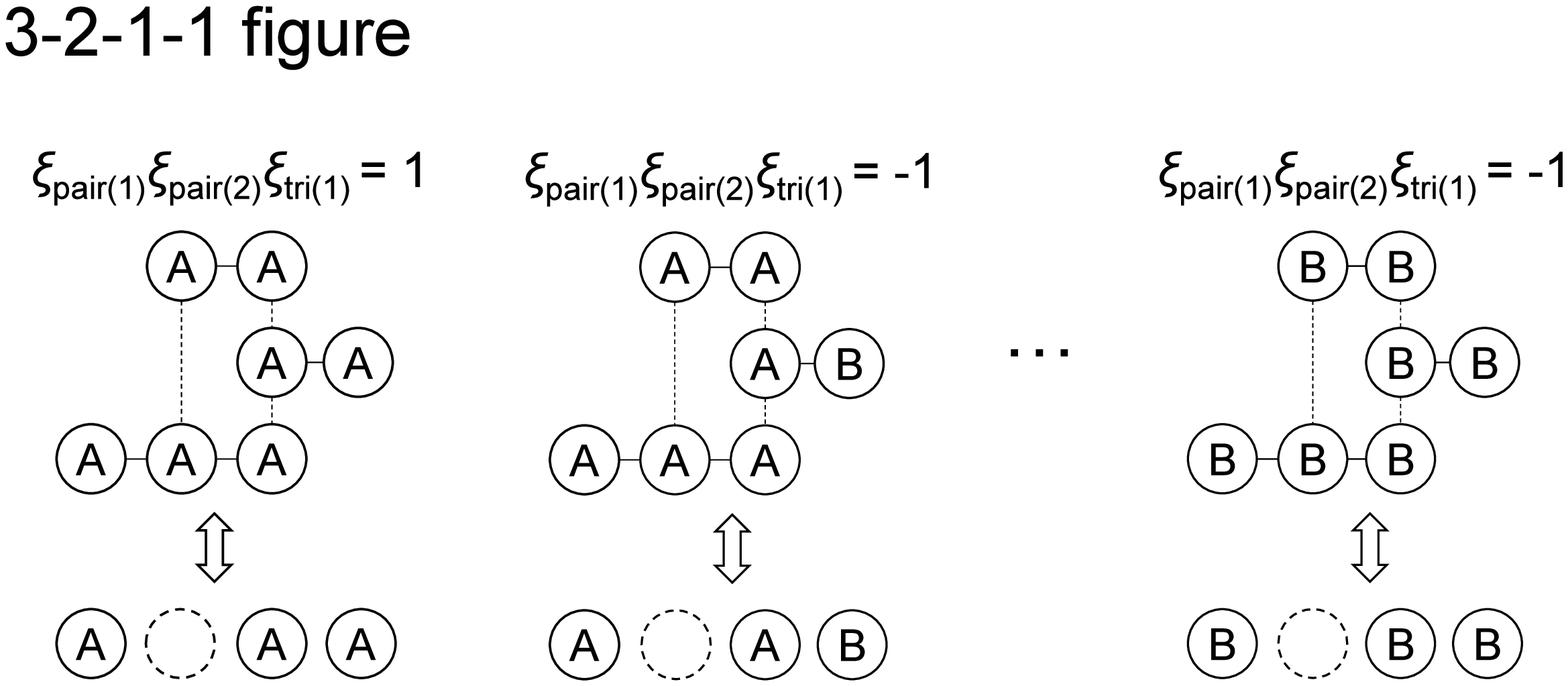}
	\caption{
		\label{fig:cluster3211}Schematic illustration that shows 3-2-1-1 figure can be considered as 1-1-1 figure.}
\end{figure}

\begin{figure}[h]
\begin{center}
\includegraphics[width=0.95\linewidth]
{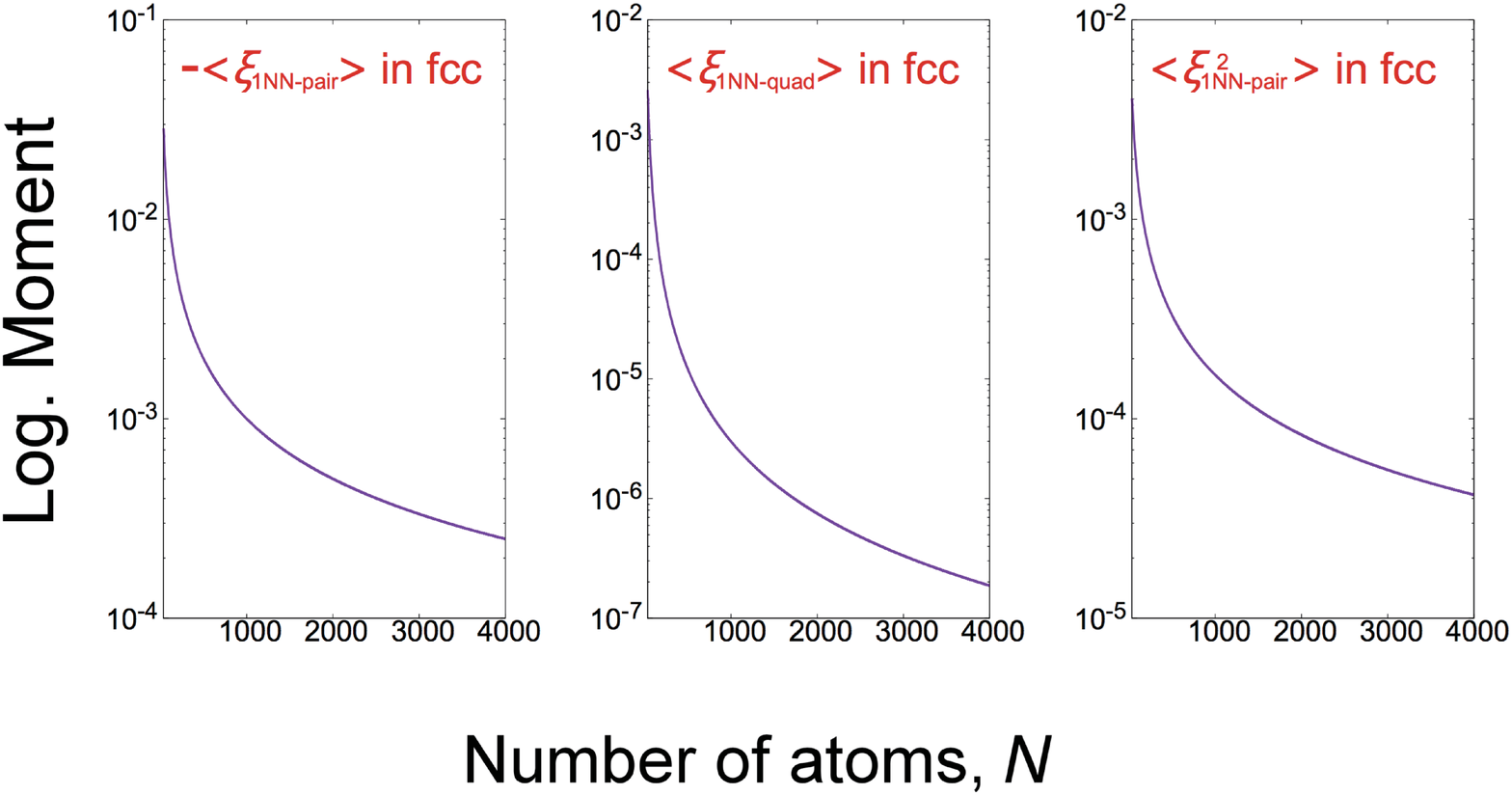}
\caption{N dependence of $ -\Braket{\xi_{\rm pair}} $,~$ \Braket{\xi_{\rm quad}} $,~$ \Braket{\xi_{\rm pair}^2} $ in fcc.}
\label{fig:moment}
\end{center}
\end{figure}

Next, according to above derivation, we actually derive the exact analytic formula about $ N $ of first order moment (i.e. average) of pair, triplet and quad cluster correlation function and second order moment (i.e. variance) of pair cluster in equiatomic fcc and bcc. From eq.~(\ref{eq:1dm-0.5-2}), the first  order moment is expressed as follow, independently of lattice, 
\begin{eqnarray}
	\Braket{\xi_{\rm pair}} &=& -\dfrac{1}{N-1},\\[10pt]
	\Braket{\xi_{\rm tri}} &=& 0,\\[8pt]
	\Braket{\xi_{\rm quad}} &=& \dfrac{3}{(N-1)(N-3)}.
\end{eqnarray} 
From Eq.~(\ref{eq:1dm-0.5-2}) and ~(\ref{eq:mdm3}) the second order moment is expressed as follow using $ F_2(4,0),~F_2(2,1),~F_2(0,2) $,
\begin{widetext}
\begin{eqnarray}
	\Braket{\xi_{\rm pair}^{2}} &=& \dfrac{1}{{}_NC_{\frac{N}{2}}\cdot N_{\rm pair}^{2}}(F_2(0,2)f_2(0,2)+F_2(2,1)f_2(2,1)+F_2(4,0)f_2(4,0))\nonumber\\
	&=&\dfrac{1}{{}_NC_{\frac{N}{2}}\cdot N_{\rm pair}^{2}}(F_2(0,2)f_1(0)+F_2(2,1)f_1(2)+F_2(4,0)f_1(4))\nonumber\\
	&=&\dfrac{1}{N_{\rm pair}^{2}}\biggl(F_2(0,2)-\dfrac{F_2(2,1)}{N-1}+\dfrac{3\cdot F_2(4,0)}{(N-1)(N-3)}\biggr).
	\label{eq:pair-mo}
\end{eqnarray}
\end{widetext}
Then, defining the coordination number about considering pair figure as $ D_{\rm pair} $, $ N_{\rm pair}$ and $ F_2(0,2) $ can be rewritten as $ND_{\rm pair}/2 $, and $ F_2(0,2) $ can be calculated as permutation of 2 from $ND_{\rm pair} $ because it is the number of combinations that two considering pair figure in different position occupy each one lattice point. Finally, calculating $ F_2(4,0) $ as number of the other combinations, Eq.~(\ref{eq:pair-mo}) can be rewritten as 
\begin{widetext}
	\begin{eqnarray}
		\Braket{\xi_{\rm pair}^{2}} = \dfrac{4}{(ND_{\rm pair})^2}\biggl(\dfrac{ND_{\rm pair}}{2}-\dfrac{N\cdot{}_{ND_{\rm pair}} {\rm P}_2}{N-1}+\dfrac{3\cdot\Bigl\{\Bigl(\dfrac{ND_{\rm pair}}{2}\Bigr)^2 -N\cdot\Bigl(\dfrac{D_{\rm pair}}{2}+{}_{ND_{\rm pair}} {\rm P}_2\Bigr) \Bigr\}}{(N-1)(N-3)}\biggr).
		\label{eq:pair-mo}
	\end{eqnarray}
\end{widetext}
The approximation formula in previous study \cite{mo} is equivalent to focusing only the first term of Eq.~(\ref{eq:pair-mo}). Table \ref{table:comvalue} is comparison of exact value in this research, approximation value in previous study and true value in the simulation in which we calculated correlation function about all atomic configuration, about first order moment of 1NN pair, triplet and quad cluster correlation function and second order moment of 1NN pair cluster in equiatomic fcc and bcc supercell , which is $2\times2\times2$ unit cell. ($D_{\rm 1NN-pair}$ is 12 in fcc and 8 in bcc.)
\begin{table}[H]
		\caption{Comparison of exact value in this research, approximation value in previous and true value in the simulation, about up to second order moment of 1NN pair cluster in equiatomic fcc and bcc supercell , which is $2\times2\times2$ unit cell.}
		\label{table:comvalue}
		\centering
		\begin{tabular}{cccc}
			\hline
			Order and lattice  & ~~~~In this study~~~~ & Approximation & ~~~~simulation~~~~ \\
			\hline \hline\\[-1pt]
			$ \Braket{\xi_{\rm pair}} $ in fcc    & $ -3.226\times10^{-2}$  & 0 & $ -3.226\times10^{-2}$\\[7.5pt]
			$ \Braket{\xi_{\rm pair}} $ in bcc    & $ -6.667\times10^{-2}$  & 0 & $ -6.667\times10^{-2}$ \\[7.5pt]
			$ \Braket{\xi_{\rm tri}} $ in fcc   & 0  & 0 & 0\\[7.5pt]
			$ \Braket{\xi_{\rm tri}} $ in bcc   & 0  & 0 & 0\\[7.5pt]
			$ \Braket{\xi_{\rm quad}} $ in fcc   & $ 3.337\times10^{-3}$  & 0 & $ 3.337\times10^{-3}$\\[7.5pt]
			$ \Braket{\xi_{\rm quad}} $ in bcc   & $ 1.538\times10^{-2}$  & 0 & $ 1.538\times10^{-2}$\\[7.5pt]
			$ \Braket{\xi_{\rm pair}^2}  $ in fcc  & $4.494\times10^{-3}$ & $5.208\times10^{-3}$ &$4.494\times10^{-3}$ \\[7.5pt]
			$ \Braket{\xi_{\rm pair}^2}  $ in bcc & $ 1.282\times10^{-2}$ & $ 1.562\times10^{-2}$ & $ 1.282\times10^{-2}$\\[2.5pt]
			\hline
		\end{tabular}
\end{table}
Figure \ref{fig:moment} shows $N$ dependence of first order moment of 1NN pair cluster correlation function and quad, and second order moment of 1NN pair in fcc.

\begin{figure}[H]
	\centering
	\includegraphics[width=0.93\linewidth]{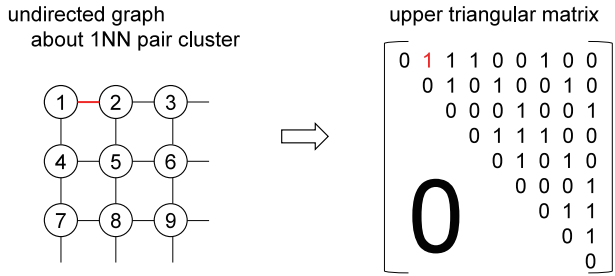}
	\caption{
		\label{fig:graph}Correspondence between undirected graph about square lattice $ 3\times3\times3 $ under periodic boundary condition and upper triangular part of adjacency matrix. The red line in the left undirected graph corresponds to the red '1' in the right upper triangular matrix.}
\end{figure}

About 3 or higher order moment, the way of counting $ F $ is exponentially complex as order of moment is higher though it can be calculated theoretically. Then, we regard lattice points and the positional relationship between those as node and edge in graph theory, that is more concretely, make upper triangular matrix from undirected graph about considering figure and lattice. (For example, Figure \ref{fig:graph} about 1NN pair in square lattice.) We can calculate $ F(p_1,~p_2,~\cdots) $ by counting the number component of tensor product of all upper triangular matrix about considering figure such that the number of number assigned to each atom appear in suffix of the component  only once is $ p_1 $ and the number of twice is $ p_2 $ and $ \cdots $.
This approach, however, needs huge amount of calculation, so we have to rely on algorithm so far. We should examine usefulness, that is, which calculation cost of this approach of calculating is lower than that of considering about all atomic configuration or not, or devise the calculation process mathematically for formulation or more simpler up to realistic range. 

\section{Conclusions}
In this study, we derived explicit expression for any-order of the generalized moment of CDOS, and confirm its validity by comparing with true value of simulation up to second order moment about one pair cluster correlation function. Additionally, we can calculate 3 or higher order moment by considering tensor product of upper triangular part of adjacency matrix for underlying graph. As we have already seen, that any-order moment can be in principle calculated, which implies that full knowledge of landscape of CDOS can be exactly determined when multiple-times tensor product can be analytically performed, which should be performed in our future study. 

\section{Acknowledgement}
This work was supported by a Grant-in-Aid for Scientific Research (16K06704) from the MEXT of Japan, Research Grant from Hitachi Metals$\cdot$Materials Science Foundation, and Advanced Low Carbon Technology Research and Development Program of the Japan Science and Technology Agency (JST).


\begin{thebibliography}{9}
\bibitem{mc1} N. Metropolis, A. W. Rosenbluth, M. N. Rosenbluth, A. H. Tellerand, and E. Teller, J. Chem. Phys. \textbf{21}, 1087 (1953).
\bibitem{mc2} A. M. Ferrenberg and R. H. Swendsen, Phys. Rev. Lett. \textbf{63}, 1195 (1989). 
\bibitem{mc3} J. Lee, Phys. Rev. Lett. 71, 211 (1993).
\bibitem{wl} F. Wang and D.P. Landau, Phys. Rev. Lett. \textbf{86}, 2050 (2001).
\bibitem{em1} K. Yuge, J. Phys. Soc. Jpn.  \textbf{84}, 084801 (2015).
\bibitem{em3} T. Taikei, T. Kishimoto, K. Takeuchi and K. Yuge, J. Phys. Soc. Jpn. \textbf{86}, 114802 (2017).
\bibitem{cm} K. Yuge, T. Taikei and K. Takeuchi, arXiv:1706.08796 [cond-mat.dis-nn].
\bibitem{em2} K. Yuge, J. Phys. Soc. Jpn. \textbf{85}, 024802  (2016).
\bibitem{ce} J.M. Sanchez, F. Ducastelle and D. Gratias, Physica A \textbf{128}, 334 (1984).
\bibitem{mo} Koretaka Yuge, Tetsuya Taikei,
and Kazuhito Takeuchi arXiv:1706.08796 [cond-mat.dis-nn]
\end{thebibliography}
\end{document}